\documentclass[11pt]{article}
\usepackage{epstopdf}
\usepackage{color}
\usepackage{subfigure}
\usepackage{amsmath}
\usepackage{amssymb}
\usepackage{graphicx,color}
\usepackage{cite}
\usepackage{enumerate}
\usepackage{amsthm}
\usepackage{amsfonts,mathrsfs}
\usepackage{geometry}
\usepackage{ifthen}

\begin{document}

\title{The $n$-th decay rate of coherence for Bell-diagonal states under quantum channels}

\vskip0.1in
\author{\small Huaijing Huang$^1$, Zhaoqi Wu$^1$\thanks{Corresponding author. E-mail: wuzhaoqi\_conquer@163.com}, Shao-Ming
Fei$^{2,3}$\\
{\small\it  1. Department of Mathematics, Nanchang University,
Nanchang 330031, P R China}\\
{\small\it  2. School of Mathematical Sciences, Capital Normal University, Beijing 100048, P R China}\\
{\small\it  3. Max-Planck-Institute for Mathematics in the Sciences,
04103 Leipzig, Germany} }

\date{}
\maketitle

\noindent {\bf Abstract} {\small } We study the degree to which the
coherence of quantum states is affected by noise. We give the
definition of the $n$-th decay rate and investigate the coherence of
Bell-diagonal states under $n$ iterations of channels. We derive
explicit formulas of the $n$-th decay rates based on $l_1$ norm of
coherence, relative entropy of coherence and skew information-based
coherence. It is found that the larger $n$ is, the faster the $n$-th
decay rate decreases as the parameter $p$ of Bell-diagonal states
increases. Moreover, for any fixed $n$, with the increase of $p$,
Bell-diagonal states can be completely incoherent under generalized
amplitude damping (GAD) channels, depolarization (DEP) channels and
phase flip (PF) channels, while this is not the case for bit flip
(BF) channels and bit-phase flip (BPF) channels. We also investigate
the geometry of the relative entropy of coherence and skew
information-based coherence of Bell-diagonal states under different
channels when the $n$-th decay rate is one, i.e., the coherence is
frozen. It is shown that compared with BF and BPF channels, when $n$
is large enough, the coherence of Bell-diagonal states will not be
frozen under GAD, DEP and PF channels. For skew information-based
coherence, similar properties of coherence freezing are found.

\vskip 0.1in


\vskip0.2in

\noindent {\bf 1. Introduction}

Originated from the quantum superposition principle, the quantum
coherence is a basic feature and a fundamental issue in quantum
mechanics and quantum information theory. The study on quantum
coherence plays an instrumental role in studying quantum
entanglement \cite{RH}, quantum correlation \cite{HO,LH}, and other
quantum phenomena \cite{JS}. The quantification of coherence
\cite{TB,SR,DG,CN,AE,MP}, the operational resource theory of
coherence \cite{AW1,EC3,EC4,EC5}, and different interpretations of
coherence \cite{CS1,JM,AU,LW,CS2} have been extensively explored
during the past few years. Quantum coherence has also been applied
in many other emergent fields, such as quantum metrology
\cite{VG1,RDD,VG2}, quantum optics \cite{RJ,EC1,LM} and quantum
biology \cite{GS,MB,EC2,SL,CML,SH}.

The geometry of measures to characterize entanglement, discord and
coherence can provide intuitions towards the quantification of these
quantities. The level surfaces of entanglement and quantum discord
for Bell-diagonal states \cite{MDL}, the level surfaces of quantum
discord for a class of two-qubit states \cite{BL}, the surfaces of
constant quantum discord and super-quantum discord for Bell-diagonal
states \cite{YKW1} have been depicted. The geometry with respect to
relative entropy of coherence and $l_1$ norm of coherence for
Bell-diagonal states have been also investigated \cite{YKW2,YKW3}.

From the perspective of available physical resources, decoherence
process occurs due to noise. The frozen of coherence is a special
process of decoherence studied in \cite{TRB1,XDY}. It is worth
pointing out that for coherence freezing, the main difficulty lies
in the fact that generally different coherence measures may yield
different orderings of coherence \cite{CLL}. At the same time, how
to quantitatively describe the decay of coherence is also an active
research topic \cite{AW2}. Recently, the decay rate of a quantum
channel for a quantum state has been defined and the decay rate of
$l_1$ norm of coherence in single-qubit system has been studied
\cite{DM}.

Instead of considering the decay rate, can we study the $n$-th decay
rate in which the coherence under channels for $n$ times is taken
into consideration? How does the $n$-th decay rate change under
different quantum channels? On the other hand, when the $n$-th decay
rate coherence is frozen by a quantum channel for $n$ times, what is
the level surfaces of the coherence for Bell-diagonal states under
the channel? In this paper, we will investigate the above problems.
The rest of the paper is structured as follows. We recall the
framework of coherence measures and introduce the concept of $n$-th
decay rate in Sec. $2$. In Sec. $3$, when Bell-diagonal states are
interfered by different noise channels for $n$ times, we study its
$n$-th decay rate with respect to three coherence measures. We
investigate the geometry of Bell-diagonal states whose initial
relative entropy of coherence and skew information-based coherence
are frozen in Sec. $4$. Some concluding remarks are given in Sec.
$5$.

\vskip0.1in

\noindent {\bf 2. The $n$-th decay rate of coherence for quantum
states under quantum channels}

\vskip0.1in

In this section, we first briefly recall the framework of
quantifying coherence proposed in \cite{TB}. Let $\mathcal{H}$ be a
$d$-dimensional Hilbert space, and $\mathcal{D(H)}$ the set of all
density operators on $\mathcal{H}$. A state and a channel are
mathematically described by a density operator (positive operator of
trace $1$) and a completely positive trace preserving (CPTP) map,
respectively \cite{MAN}. For a prefixed orthonormal basis
$\{|k\rangle\}^d_{k=1}$ of $\mathcal{H}$, the density operators
which are diagonal in this basis are called incoherent states.
Otherwise, they are said to be coherent. The set of incoherent
states is denoted by $\mathcal{I}$, i.e.,
$$
\mathcal{I}=\{\delta\in \mathcal{D(H)}|\delta=\sum_{k}p_k|k\rangle\langle k|,~p_k\geq 0,~\sum_{k}p_k=1\}.
$$

Let $\Lambda$ be a CPTP map,
$$\Lambda(\rho)=\sum_{i}K_i\rho K_i^\dag,$$
where $K_i$ are Kraus operators satisfying $\sum_{i}K_i^\dag
K_i=I_{d}$ with $I_d$ being the identity operator. $K_i$ are called
incoherent Kraus operators if $K_i^\dag \mathcal{I}K_i\in
\mathcal{I}$ for all $i$, and in this case the corresponding
$\Lambda$ is called an incoherent operation.

In \cite{TB}, the authors proposed a framework for a measure of coherence

(i) (Faithfulness) $C(\rho)\geq 0$ and $C(\rho)=0$ iff $\rho$ is
incoherent.

(ii) (Monotonicity) $C(\Lambda(\rho))\leq C(\rho)$ for any
incoherent operation $\Lambda$.

(iii) (Convexity) $C(\cdot)$ is a convex function of $\rho$, i.e.,
$$\sum_{n}p_nC(\rho_n)\geq C(\sum_{n}p_n\rho_n),$$
where $p_n\geq 0, \sum_{n}p_n=1$.

(iv) (Strong monotonicity) $C(\cdot)$ does not increase on average
under selective incoherent operations, i.e.,
$$C(\rho)\geq \sum_{n}p_nC(\varrho_n),$$
where $p_n=\mathrm{Tr}(K_n\rho K_n^\dag)$ are probabilities and
$\varrho_n=\frac{K_n\rho K_n^\dag}{p_n}$ are post-measurement states
with $K_n$ being incoherent Kraus operators.

The $l_{1}$ norm of coherence of a quantum state $\rho$ is given by
\cite{TB}
\begin{equation}\label{eq1}
C_{l_1}(\rho )=\min_{\delta\in
\mathcal{I}}\parallel\rho-\delta\parallel_{l_{1}}=\sum _{i\neq
j}\left| \rho _{ij}\right|,
\end{equation}
the minimal distance between $\rho$ and incoherent quantum state
$\delta$. The relative entropy of coherence measure is defined by
the following formula \cite{TB}
\begin{equation}\label{eq2}
C_r(\rho )=\min_{\delta\in
\mathcal{I}}S\left(\rho\parallel\delta\right)=S\left(\rho
_{\text{diag}}\right)-S(\rho ),
\end{equation}
where $S(\rho\parallel\delta)=\text{Tr}\rho
\text{log}\rho-\text{Tr}\rho \text{log}\delta$ is the von Neumann
entropy. The $l_{1}$ norm of coherence and relative entropy of
coherence both satisfy all of the above-mentioned conditions.

For the fixed orthonormal basis $\{|k\rangle\}^d_{k=1}$, the skew
information-based coherence is presented in \cite{CS3},
\begin{equation}\label{eq3}
C_I(\rho)=\sum_{k=1}^d I(\rho,|k\rangle\langle
k|)=1-\sum_{k=1}^d\langle k|\sqrt{\rho}|k\rangle ^2,
\end{equation}
where $I(\rho,|k\rangle\langle
k|)=-\frac{1}{2}\mathrm{Tr}\{[\sqrt{\rho },|k\rangle \langle
k|]\}^2$. It is confirmed that skew information-based coherence also
fulfills the conditions (i)-(iv) \cite{CS3}.

We are now in a position to give the definition of the $n$-th decay
rate.

{\bf Definition 1 ($n$-th decay rate of coherence).} For a quantum
channel $\Phi$ and a coherent state $\rho$, the $n$-th dacay rate is
defined as
\begin{equation}\label{eq4}
R_n(\rho^\Phi)=\frac{C \left(\Phi^n (\rho)\right)}{C (\rho)},
\end{equation}
where $n$ is the number of times that the quantum state $\rho$
passes through the channel.

From the definition, one sees that the $n$-th decay rate quantifies the decrease
of coherence on average under the channels. Now we focus on the decay
phenomenon for the following channels for a bipartite quantum state $\rho$,
$$
\Phi (\rho )=\sum _{i,j} \left(E_i\otimes E_j\right)\rho
\left(E_i\otimes E_j\right)^{\dagger},
$$
where $\left\{E_i\right\}$ are Kraus operators satifying $\sum_i
E_i^{\dagger}E_i=I_{d}$. Denote $\Phi^1(\rho)=\Phi(\rho)$. For
$1<n\in\mathbb{N}$, we define the $n$-th iteration of a quantum
channel by
$$
\Phi ^n(\rho )=\sum _{i,j} \left(E_i\otimes E_j\right)\Phi
^{n-1}(\rho )\left(E_i\otimes E_j\right)^{\dagger}.
$$
When $R_n(\rho^\Phi)=1$, the coherence keeps invariant and the
initial coherence is frozen by the channel $\Phi$ \cite{TRB1}.

\vskip0.1in

\noindent {\bf 3. Geometry of the $n$-th decay rates of coherence
for Bell-diagonal states under $n$ iterations of quantum channels}

\vskip0.1in

We now calculate the $n$-th decay rates of coherence for
Bell-diagonal states under Markovian channels. Two-qubit
Bell-diagonal states can be expressed as
\begin{equation}\label{eq5}
\rho =\frac{1}{4}\left(I\otimes I+\sum _{i=1}^3 c_i\sigma _i\otimes
\sigma _i\right),
\end{equation}
where $\left\{\sigma _i\right\}_{i=1}^3$ are the standard Pauli
matrices and $c_1,c_2,c_3\in [-1,1]$. In the computational basis
$\{|00\rangle ,|01\rangle ,|10\rangle ,|11\rangle \}$, the matrix
form of $\rho$ can be written as
$$
\rho=\frac{1}{4}\left(
\begin{array}{cccc}
 1+c_3 & 0 & 0 & c_1-c_2 \\
 0 & 1-c_3 & c_1+c_2 & 0 \\
 0 & c_1+c_2 & 1-c_3 & 0 \\
 c_1-c_2 & 0 & 0 & 1+c_3 \\
\end{array}
\right).
$$
{\begin{table} \caption{  {Kraus operators for the quantum channels:
bit flip (BF), phase flip (PF), bit-phase flip (BPF), depolarizing
(DEP), and generalized amplitude damping (GAD), where $p$ and
$\gamma$ are decoherence probabilities, $0<p<1,~0<\gamma<1$.} }
$$\begin{array}{cccccc}\hline\hline
\text{Channel}\hspace{1.3in}   &  \text{Kraus operators}\\ \hline
 \mathrm{BF}\hspace{1.3in}   &   E_0=\sqrt{1-p/2}I,\,\,\,\,\,\,  E_1=\sqrt{p/2}\sigma_1  \\
 \mathrm{PF}\hspace{1.3in}   &   E_0=\sqrt{1-p/2}I,\,\,\,\,\,\,  E_1=\sqrt{p/2}\sigma_3  \\
 \mathrm{BPF}\hspace{1.3in}  &   E_0=\sqrt{1-p/2}I,\,\,\,\,\,\,  E_1=\sqrt{p/2}\sigma_2  \\
 \mathrm{DEP}\hspace{1.3in}  &   E_0=\sqrt{1-p}I,\,\,\,\,\,\,  E_1=\sqrt{p/3}\sigma_1    \\
                             &   E_2=\sqrt{p/3}\sigma_2,\,\,\,\,\,\, E_3=\sqrt{p/3}\sigma_3   \\
 \mathrm{GAD}\hspace{1.3in}  &   E_0=\sqrt{p}\left(\begin{array}{cc}
         1&0\\
         0&\sqrt{1-\gamma}\\
         \end{array}
         \right), \,\,\,\,\,\, E_2=\sqrt{1-p}\left(\begin{array}{cc}
         \sqrt{1-\gamma}&0\\
         0&1\\
         \end{array}
         \right)  \\
  &  E_1=\sqrt{p}\left(\begin{array}{cc}
         0&\sqrt{\gamma}\\
         0&0\\
         \end{array}
         \right), \,\,\,\,\,\, E_3=\sqrt{1-p}\left(\begin{array}{cc}
         0&0\\
         \sqrt{\gamma}&0\\
         \end{array}
         \right)\\ \hline\hline
\end{array}$$
\end{table}}

We consider five quantum channels whose Kraus operators are listed
in Table 1 \cite{YKW1}. Note that the Bell-diagonal states remain in
the same form under these channels. When the Bell-diagonal states
are subjected to $n$ iterations of these five channels, the
coefficients $c_i(i=1,2,3)$  will be transformed to
$c_i^{*}(i=1,2,3)$, see Table 2 \cite{YKW2}.

{\begin{table} \caption{ {Correlation coefficients of Bell-diagonal
states under $n$ iterations of five channels: bit flip
$(\text{BF}^{n})$, phase flip $(\text{PF}^{n})$, bit-phase flip
$(\text{BPF}^{n})$, depolarizing $(\text{DEP}^{n})$ and generalized
amplitude damping $(\text{GAD}^{n})$. For the coefficients of
$\text{GAD}^{n}$ in the last row, we have fixed $p=1/2$ and replaced
$\gamma$ by $p$ in the Kraus operators $E_i(i=0,1,2,3)$ mentioned in
Table 1.} }
$$\begin{array}{cccccc}\hline\hline
\text{Channel}\hspace{0.9in} & \text{$c_1^{*}$} \hspace{0.9in} & \text{$c_2^{*}$} \hspace{0.9in} & \text{$c_3^{*}$}\\
\hline
 \mathrm{BF}^n\hspace{0.9in}  &  c_1        \hspace{0.9in}  & c_2(1-p)^{2n}  \hspace{0.9in} &  c_3(1-p)^{2n}\\
 \mathrm{PF}^n\hspace{0.9in}  &  c_1(1-p)^{2n} \hspace{0.9in}  & c_2(1-p)^{2n}  \hspace{0.9in} &  c_3\\
 \mathrm{BPF}^n\hspace{0.9in} &  c_1(1-p)^{2n} \hspace{0.9in}  & c_2         \hspace{0.9in} &  c_3(1-p)^{2n}\\
 \mathrm{DEP}^n\hspace{0.9in} &  c_1(1-\frac{4}{3}p)^n   \hspace{0.9in}  & c_2(1-\frac{4}{3}p)^n    \hspace{0.9in} &  c_3(1-\frac{4}{3}p)^n\\
 \mathrm{GAD}^n\hspace{0.9in} &  c_1(1-p)^n   \hspace{0.9in}  & c_2(1-p)^n    \hspace{0.9in} &  c_3(1-p)^{2n}\\ \hline\hline
\end{array}$$
\end{table}}

From (\ref{eq4}) we can calculate the $n$-th decay rates of $l_{1}$
norm of coherence, relative entropy of coherence and skew
information-based coherence for Bell-diagonal states, respectively,
\begin{equation*}
 R_n^l(\rho^\Phi)=\frac{\left| c_1^*-c_2^*\right| +\left|c_1^*+c_2^*\right| }{\left| c_1-c_2\right| +\left| c_1+c_2\right| },
\end{equation*}
\begin{equation*}
\aligned
R_n^r(\rho^\Phi)&=[(1-c_1^{*}+c_2^{*}-c_3^{*})\ln(1-c_1^{*}+c_2^{*}-c_3^{*})+(1+c_1^{*}-c_2^{*}+c_3^{*})\ln(1+c_1^{*}-c_2^{*}+c_3^{*})\\
&\quad+(1-c_1^{*}-c_2^{*}-c_3^{*})\ln(1-c_1^{*}-c_2^{*}-c_3^{*})+(1+c_1^{*}+c_2^{*}+c_3^{*})\ln(1+c_1^{*}+c_2^{*}+c_3^{*})\\
&\quad-2(1-c_3^{*})\ln(1-c_3^{*})-2(1+c_3^{*})\ln(1+c_3^{*})]/\\
&\quad[(1-c_1+c_2-c_3)\ln(1-c_1+c_2-c_3)+(1+c_1-c_2+c_3)\ln(1+c_1-c_2+c_3)\\
&\quad+(1-c_1-c_2-c_3)\ln(1-c_1-c_2-c_3)+(1+c_1+c_2+c_3)\ln(1+c_1+c_2+c_3)\\
&\quad-2(1-c_3)\ln(1-c_3)-2(1+c_3)\ln(1+c_3)],
\endaligned
\end{equation*}
and
\begin{equation*}
\aligned
R_n^I(\rho^\Phi)&=(2-\sqrt{1-c_1^*-c_2^*-c_3^*}\sqrt{1+c_1^*+c_2^*-c_3^*}-\sqrt{1+c_1^*-c_2^*+c_3^*}\sqrt{1-c_1^*+c_2^*+c_3^*})\\
&\quad/(2-\sqrt{1-c_1-c_2-c_3}
\sqrt{1+c_1+c_2-c_3}-\sqrt{1+c_1-c_2+c_3} \sqrt{1-c_1+c_2+c_3}).
\endaligned
\end{equation*}
\begin{figure}\centering
\subfigure[] {\begin{minipage}[b]{0.3\linewidth}
\includegraphics[width=1\textwidth]{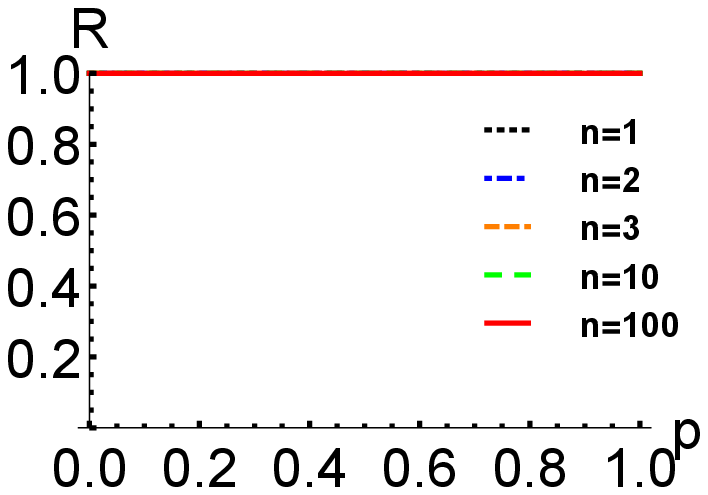}
\end{minipage}}
\subfigure[] {\begin{minipage}[b]{0.3\linewidth}
\includegraphics[width=1\textwidth]{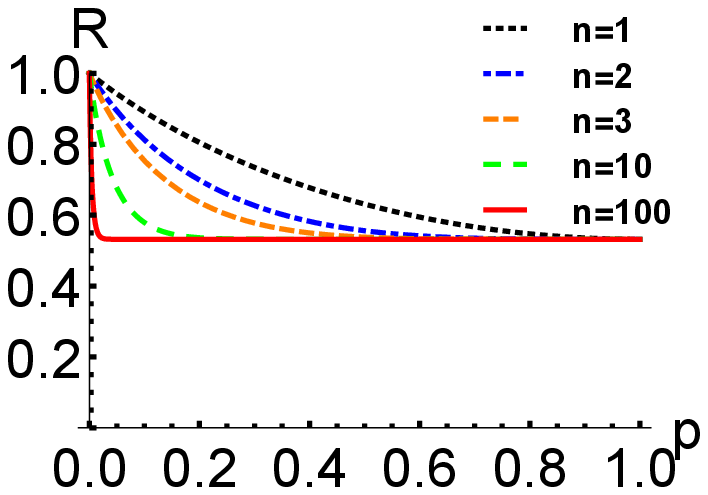}
\end{minipage}}
\subfigure[] {\begin{minipage}[b]{0.3\linewidth}
\includegraphics[width=1\textwidth]{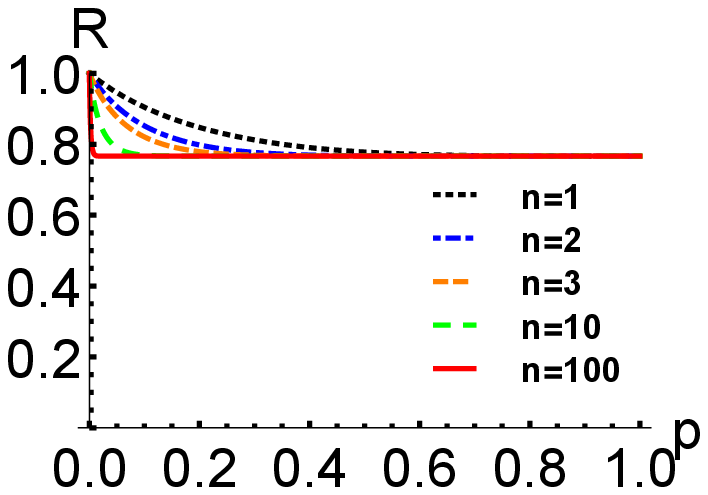}
\end{minipage}}
\caption{The $n$-th decay rates for Bell-diagonal states ($c_1=0.6,
c_2=0.1, c_3=0.2$) under $n$ iterations of bit flip channels
$\mathrm{BF}^n$: $(a)$ $l_{1}$ norm of coherence; $(b)$ relative
entropy of coherence; $(c)$ skew information-based coherence.}
\label{fig:DR1}
\end{figure}

\begin{figure}\centering
\subfigure[] {\begin{minipage}[b]{0.3\linewidth}
\includegraphics[width=1\textwidth]{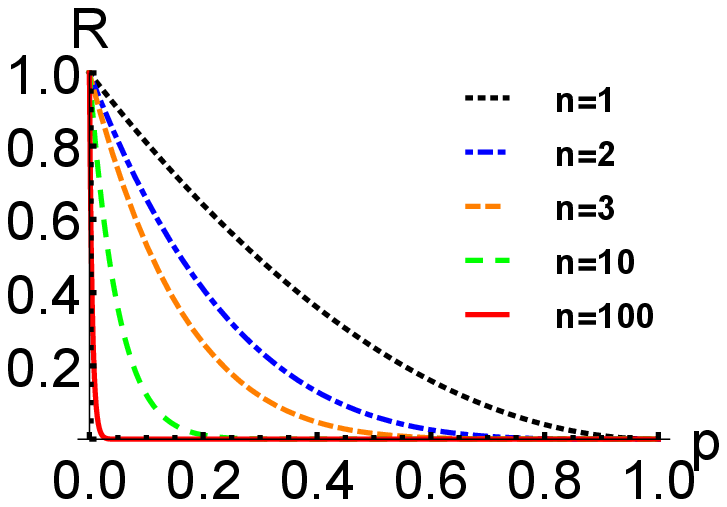}
\end{minipage}}
\subfigure[] {\begin{minipage}[b]{0.3\linewidth}
\includegraphics[width=1\textwidth]{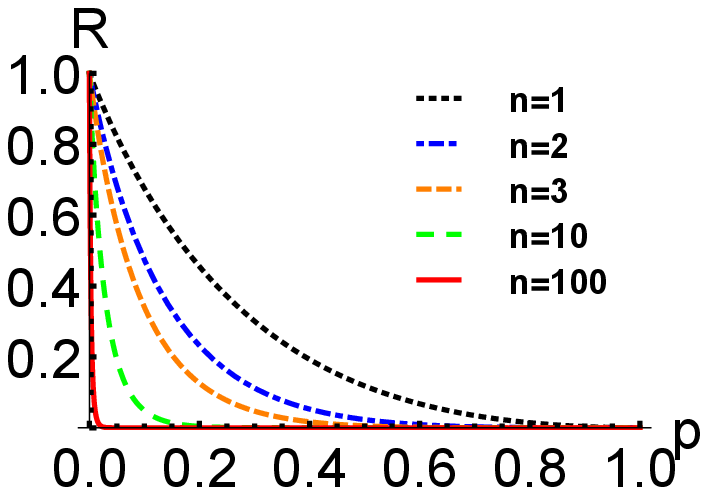}
\end{minipage}}
\subfigure[] {\begin{minipage}[b]{0.3\linewidth}
\includegraphics[width=1\textwidth]{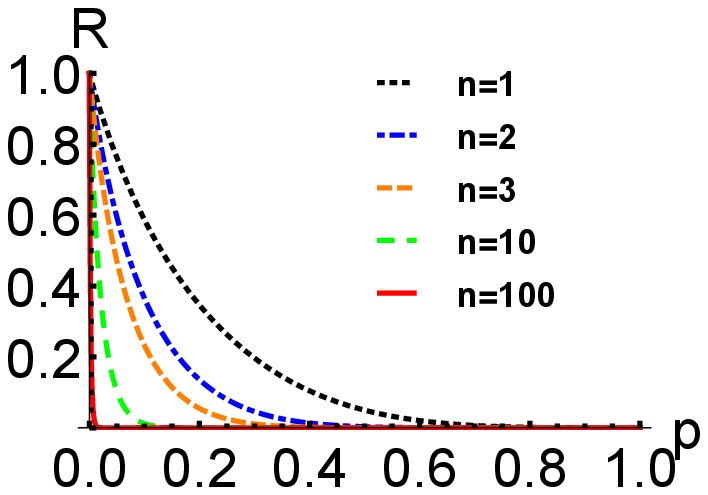}
\end{minipage}}
\caption{The $n$-th decay rates for Bell-diagonal states ($c_1=0.6,
c_2=0.1, c_3=0.2$) under $n$ iterations of phase flip channels
$\mathrm{PF}^n$: $(a)$ $l_{1}$ norm of coherence; $(b)$ relative
entropy of coherence; $(c)$ skew information-based coherence.}
\label{fig:DR2}
\end{figure}

\begin{figure}\centering
\subfigure[] {\begin{minipage}[b]{0.3\linewidth}
\includegraphics[width=1\textwidth]{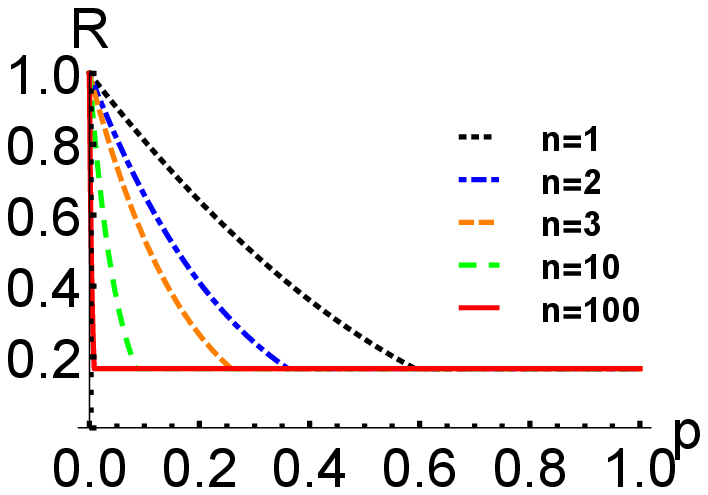}
\end{minipage}}
\subfigure[] {\begin{minipage}[b]{0.3\linewidth}
\includegraphics[width=1\textwidth]{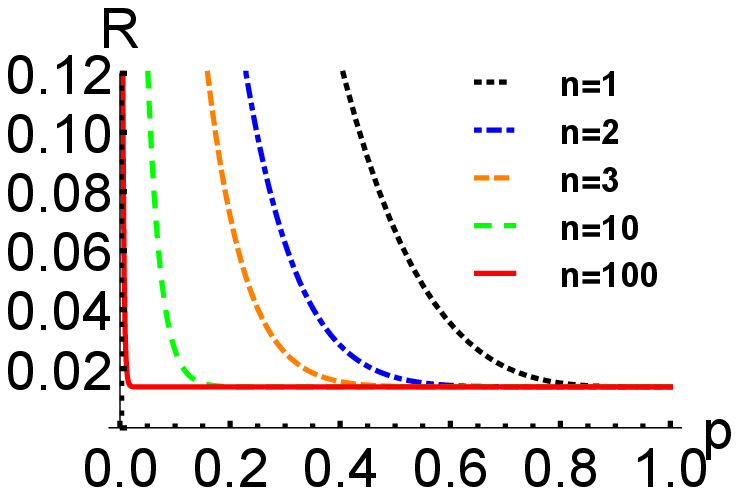}
\end{minipage}}
\subfigure[] {\begin{minipage}[b]{0.3\linewidth}
\includegraphics[width=1\textwidth]{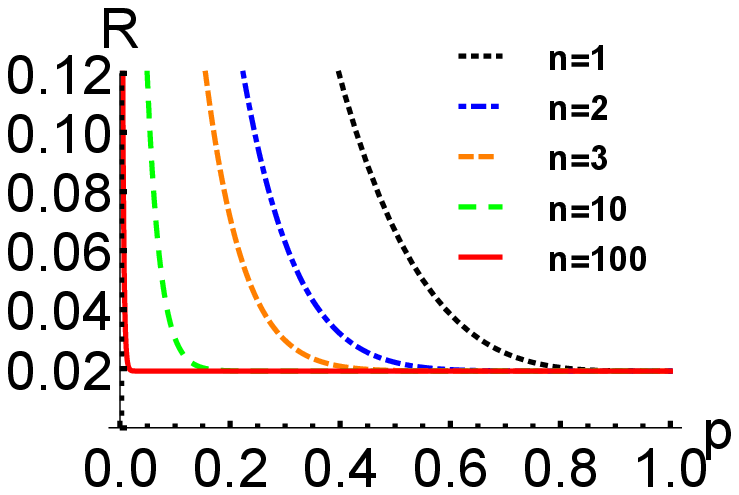}
\end{minipage}}
\caption{The $n$-th decay rates for Bell-diagonal states ($c_1=0.6,
c_2=0.1, c_3=0.2$) under $n$ iterations of bit-phase flip channels
$\mathrm{BPF}^n$: $(a)$ $l_{1}$ norm of coherence; $(b)$ relative
entropy of coherence; $(c)$ skew information-based coherence.}
\label{fig:DR3}
\end{figure}

\begin{figure}\centering
\subfigure[] {\begin{minipage}[b]{0.3\linewidth}
\includegraphics[width=1\textwidth]{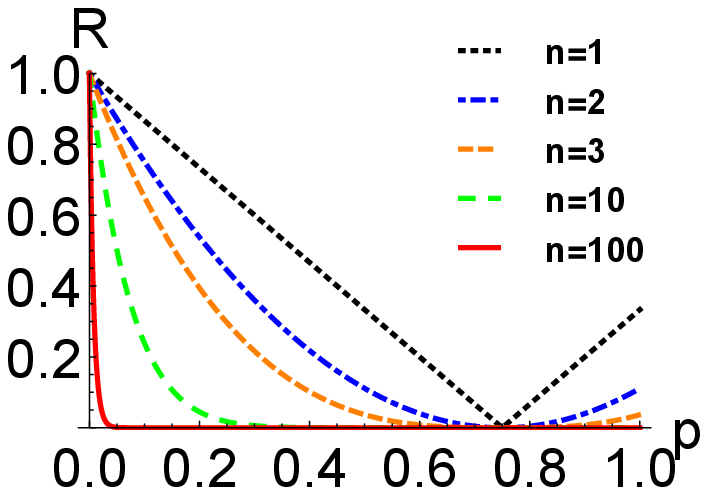}
\end{minipage}}
\subfigure[] {\begin{minipage}[b]{0.3\linewidth}
\includegraphics[width=1\textwidth]{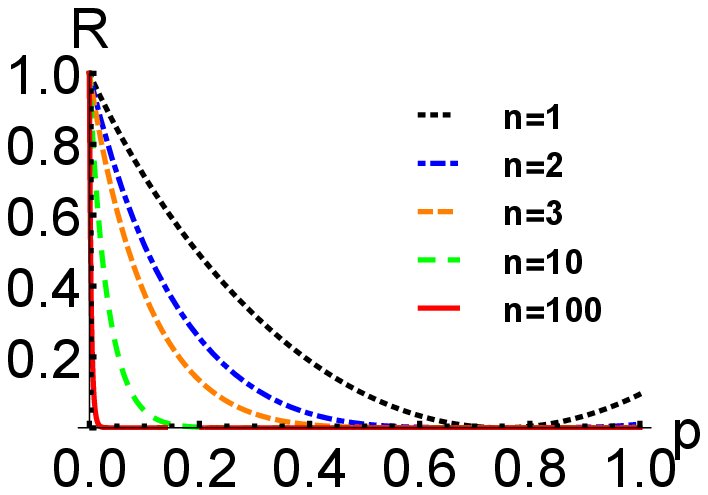}
\end{minipage}}
\subfigure[] {\begin{minipage}[b]{0.3\linewidth}
\includegraphics[width=1\textwidth]{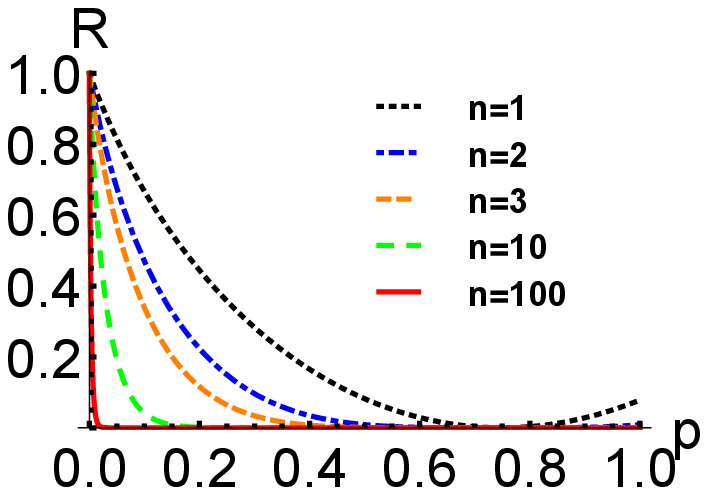}
\end{minipage}}
\caption{The $n$-th decay rates for Bell-diagonal states ($c_1=0.6,
c_2=0.1, c_3=0.2$) under $n$ iterations of depolorizing channels
$\mathrm{DEP}^n$: $(a)$ $l_{1}$ norm of coherence; $(b)$ relative
entropy of coherence; $(c)$ skew information-based coherence.}
\label{fig:DR4}
\end{figure}

\begin{figure}\centering
\subfigure[] {\begin{minipage}[b]{0.3\linewidth}
\includegraphics[width=1\textwidth]{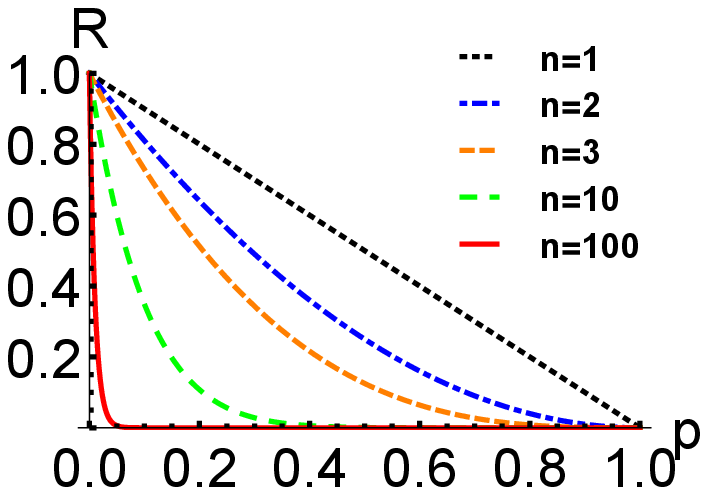}
\end{minipage}}
\subfigure[] {\begin{minipage}[b]{0.3\linewidth}
\includegraphics[width=1\textwidth]{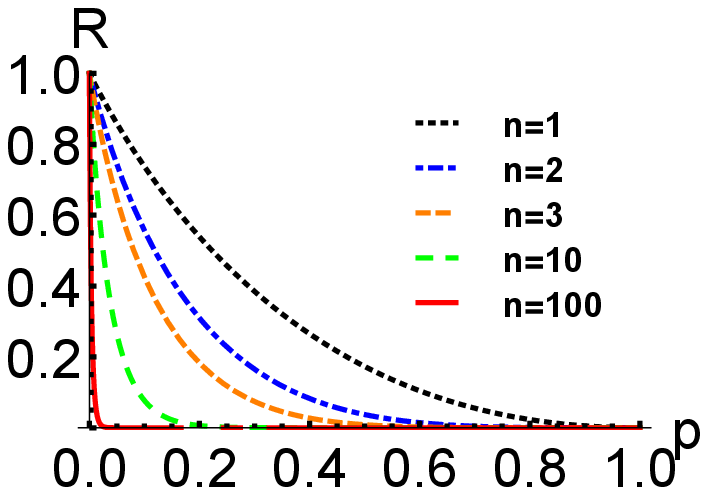}
\end{minipage}}
\subfigure[] {\begin{minipage}[b]{0.3\linewidth}
\includegraphics[width=1\textwidth]{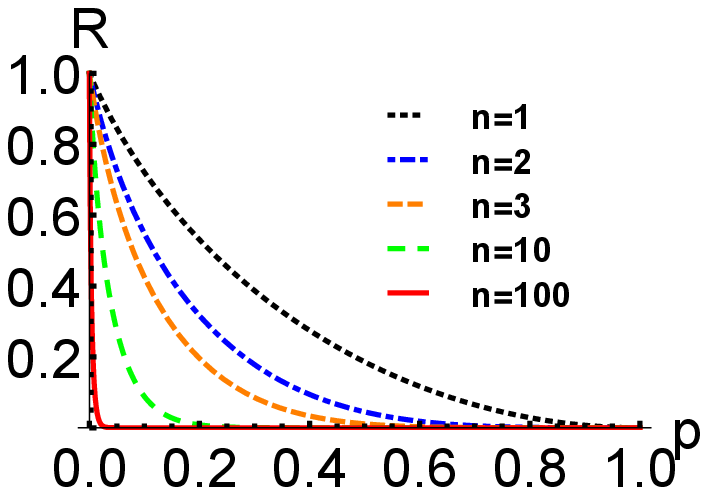}
\end{minipage}}
\caption{The $n$-th decay rates for Bell-diagonal states ($c_1=0.6,
c_2=0.1, c_3=0.2$) under $n$ iterations of generalized amplitude
damping channels $\mathrm{GAD}^n$: $(a)$ $l_{1}$ norm of coherence;
$(b)$ relative entropy of coherence; $(c)$ skew information-based
coherence.} \label{fig:DR5}
\end{figure}

Note that for DEP channels, we get $R_n(\rho^\Phi)=0$ if
$p=\frac{3}{4}$, which indicates that these channels make state
$\rho$ completely incoherent. The coherence of Bell-diagonal states
is generally different under different types of coherence measures.
The influence of channels on coherence, for Bell-diagonal state with
parameters $c_1=0.6, c_2=0.1$ and $c_3=0.2$, are shown in Figures 1,
2, 3, 4 and 5. From these figures, it is observed that for different
coherence measures and the same channels, the trend of the coherence
of Bell-diagonal states is the same, i.e., the $n$-th decay rates
decrease for any $n$. For different channels, except for BF, the
decoherence process of Bell-diagonal states is very similar under
relative entropy of coherence and skew information-based coherence.
Compared with relative entropy of coherence and skew
information-based coherence, the curvature of the decay rate curve
of Bell-diagonal states is obviously different from that of $l_{1}$
norm of coherence. It is found that the larger $n$ is, the faster
the $n$-th decay rates decrease as $p$ increases. The $n$-th decay
rates decrease with the increase of $p$ for fixed $n$. When the
$n$-th decay rates goes to 0, channels have the greatest influence
on coherence.

The $n$-th decay rates of coherence for Bell-diagonal states under
PF, DEP and GAD channels could reach 0, as displayed in Figures 2, 4
and 5. This implies that for some particular $p$, the coherence of
the Bell-diagonal state  vanishes under the quantum channel.
However, for BF and BPF channels, the $n$-th decay rates reach a
certain value greater than 0, regardless of $n$, see Figures 1 and
3, namely, they have freezing effect on coherence. Specifically,
under BF channels, the $l_{1}$ norm of coherence of Bell-diagonal
states is frozen, see Figure 1$(a)$. Comparing it with Figures
1$(b)$ and 1$(c)$, we find that the $n$-th decay rates of relative
entropy of coherence and skew information-based coherence for
Bell-diagonal states firstly decrease under BF channels, and then
keep unchanged. That is, the decoherence process occurs first, and
then the coherence freezing phenomenon appears. For DEP channels,
the $n$-th decay rates reduce to zero firstly and then increases,
see Figure 4.

\vskip0.1in

\noindent {\bf 4. Geometry of frozen coherence for Bell-diagonal
states under $n$ iterations of quantum channels}

\vskip0.1in

In this section, we mainly study the geometry of relative entropy of
coherence and skew information-based coherence for Bell-diagonal
states under different channels when the coherence of initial states
is frozen under $n$ iterations of the quantum channel $\Phi$, i.e.,
$R_n(\rho^\Phi)=1$. We investigate the geometry of Bell-diagonal
states for relative entropy of coherence. The surfaces of coherence
freezing for Bell-diagonal states under different channels for fixed
$p=0.5$ and $n$ are shown in Figures 6, 7, 8, 9 and 10. It is found
that different channels have different conditions which make the
coherence of Bell-diagonal states frozen.

Now fix $p=0.5$. For BF and BPF channels, when the coherence of
Bell-diagonal states is frozen, with the increase of $n$, the
coherence curve gradually evolves from $(a)$ to $(c)$, see Figures 6
and 8. Moreover, we discover that with the increase of $p$, the
coherence surfaces will reach the state of $(c)$ faster for fixed
$n$. For both channels, obvious changes happen when $n$ varies from
$11$ to $12$.
\begin{figure}[htbp]\centering
\subfigure[] {\begin{minipage}[b]{0.3\linewidth}
\includegraphics[width=1\textwidth]{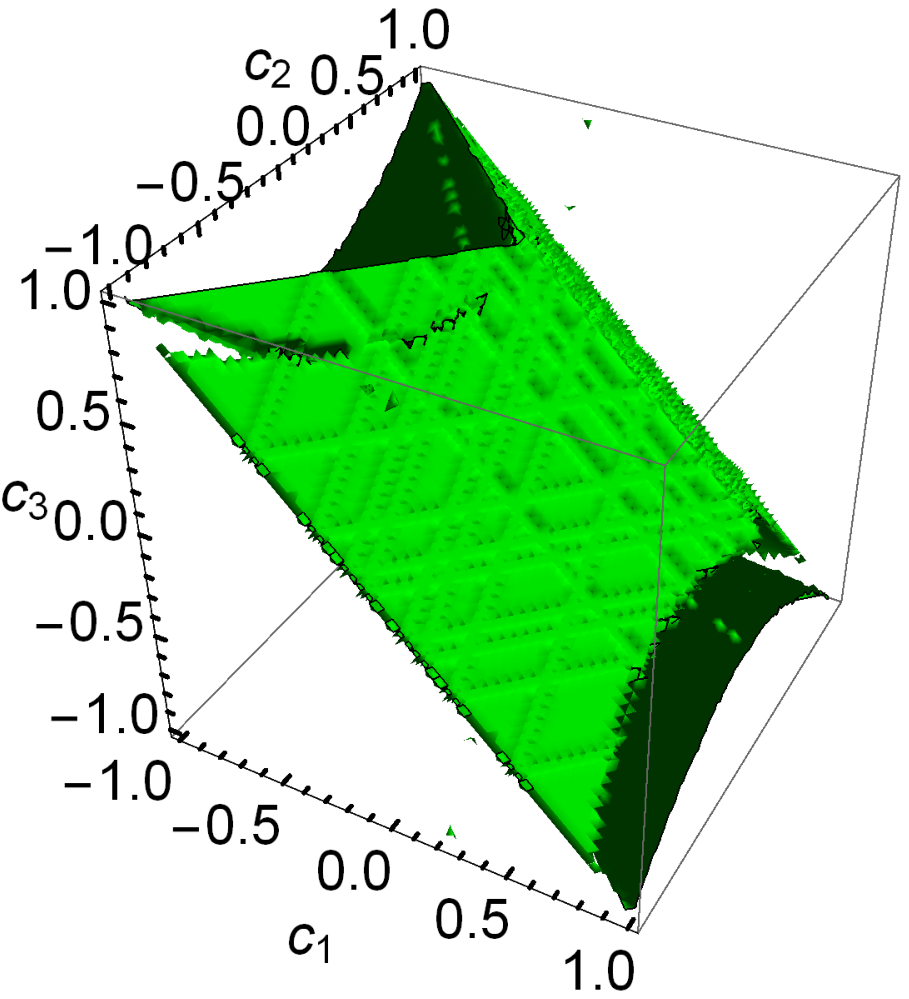}
\end{minipage}}
\subfigure[] {\begin{minipage}[b]{0.3\linewidth}
\includegraphics[width=1\textwidth]{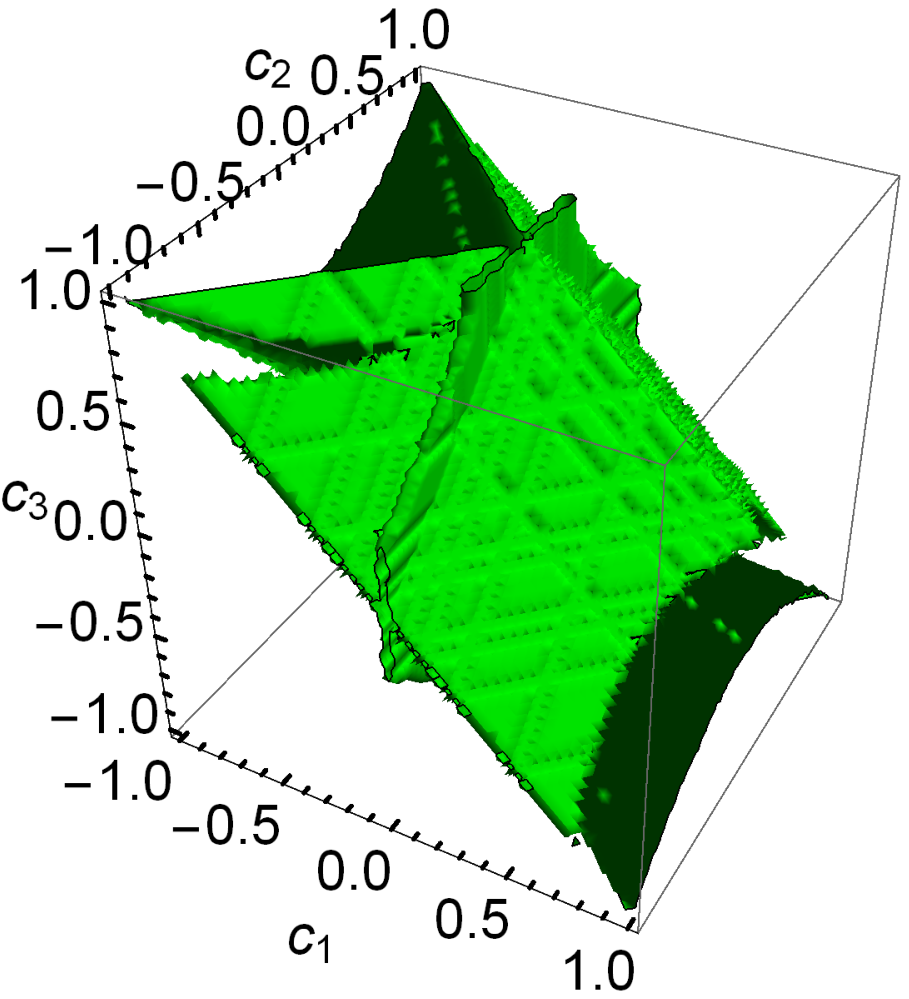}
\end{minipage}}
\subfigure[] {\begin{minipage}[b]{0.3\linewidth}
\includegraphics[width=1\textwidth]{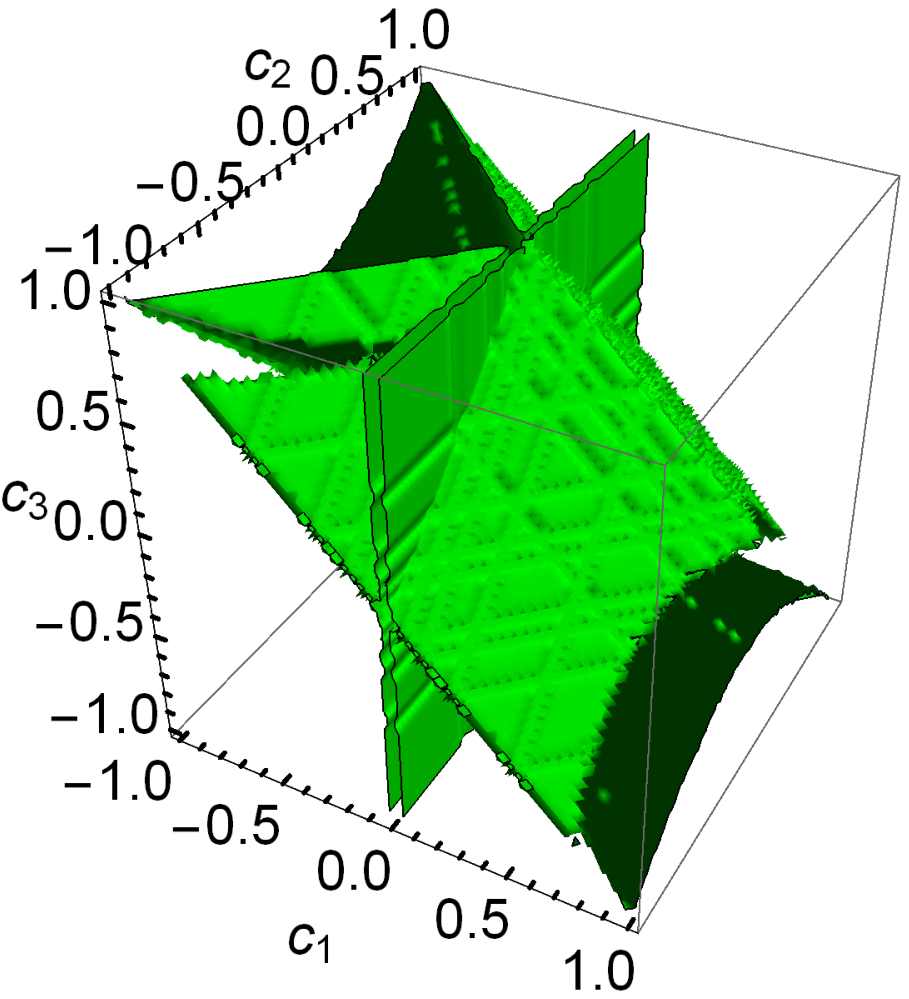}
\end{minipage}}
\caption{Surfaces of relative entropy of coherence freezing for
Bell-diagonal states under $n$ iterations of bit flip channels
$\mathrm{BF}^n$: $(a)$ $p=0.5$, $n=1$; $(b)$ $p=0.5$, $n=11$; $(c)$
$p=0.5$, $n=12$.} \label{fig:DR6}
\end{figure}

\begin{figure}[htbp]\centering
\subfigure[] {\begin{minipage}[b]{0.3\linewidth}
\includegraphics[width=1\textwidth]{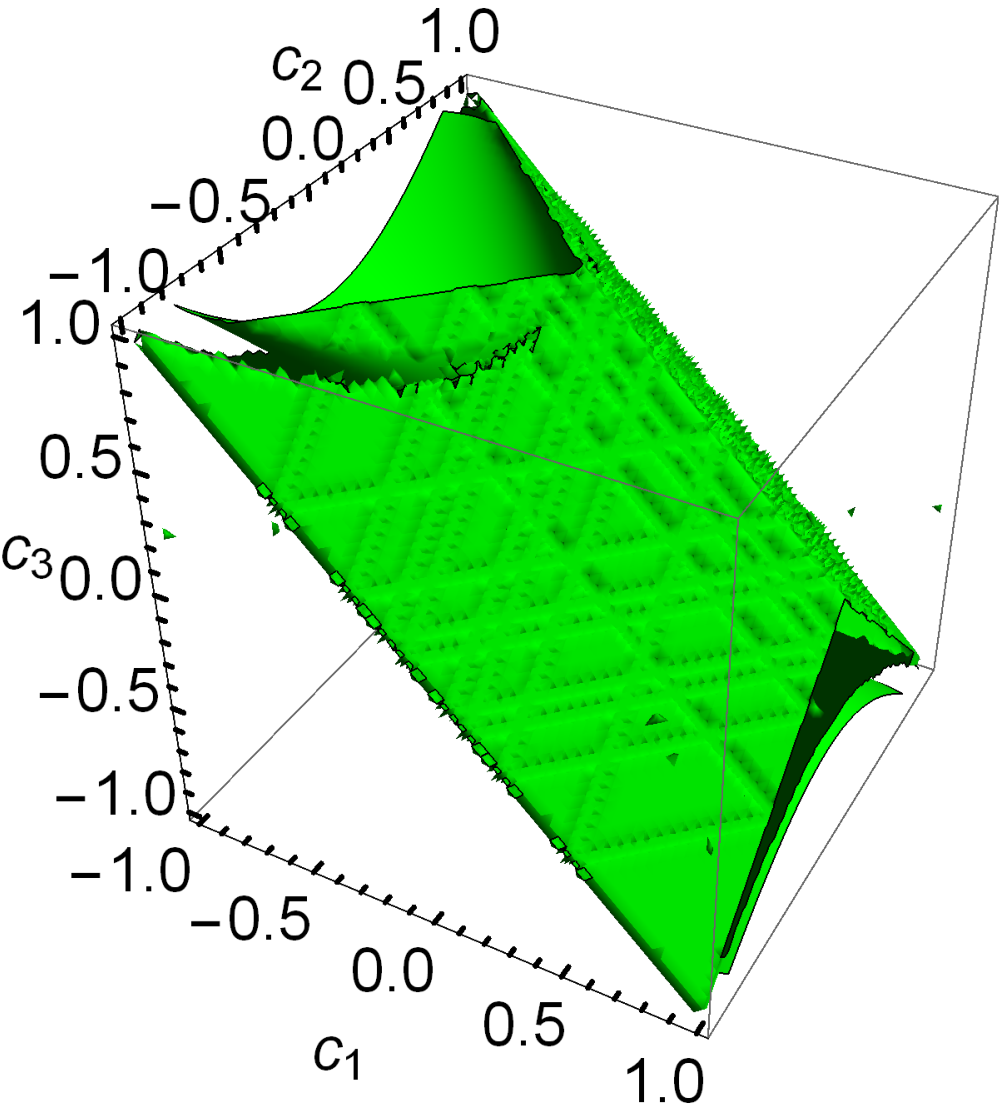}
\end{minipage}}
\subfigure[] {\begin{minipage}[b]{0.3\linewidth}
\includegraphics[width=1\textwidth]{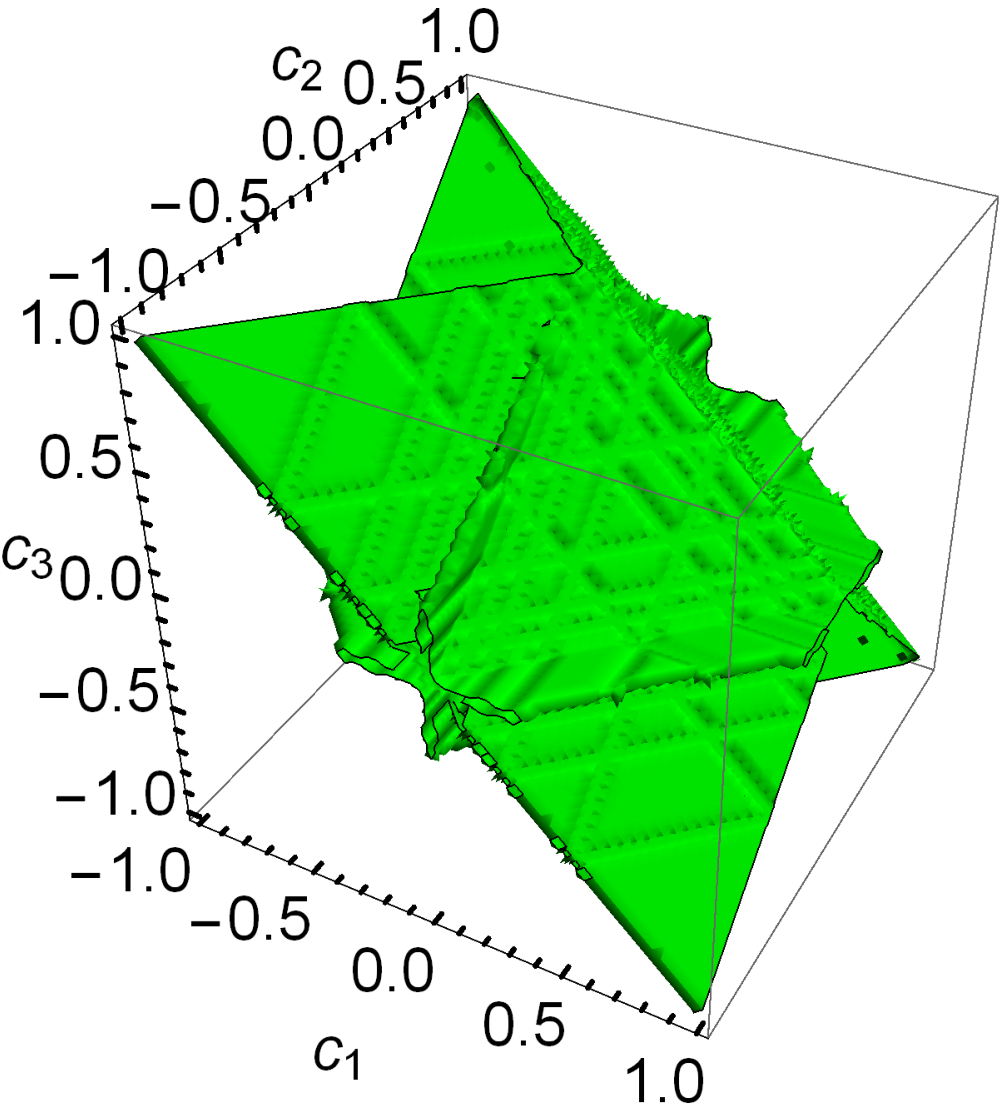}
\end{minipage}}
\subfigure[] {\begin{minipage}[b]{0.3\linewidth}
\includegraphics[width=1\textwidth]{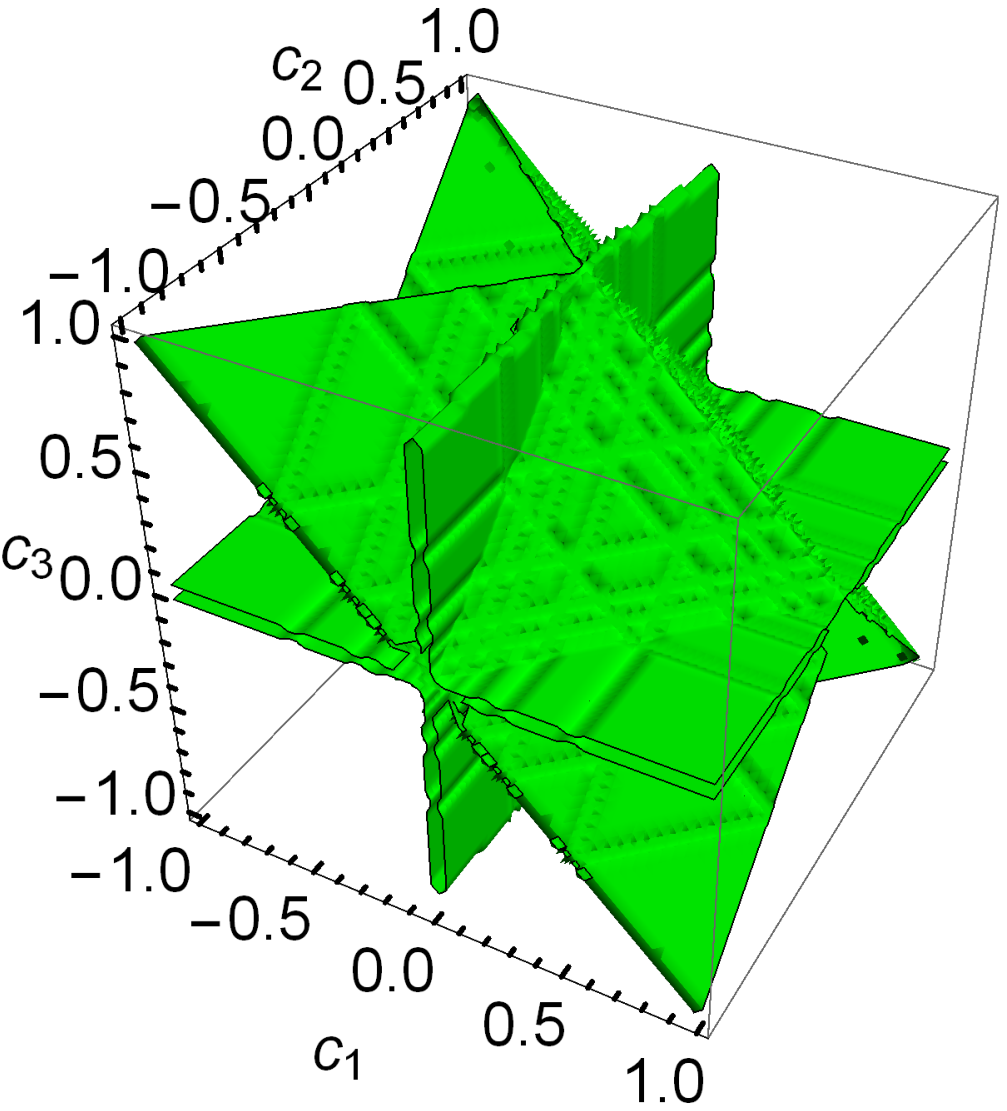}
\end{minipage}}
\caption{Surfaces of relative entropy of coherence freezing for
Bell-diagonal states under $n$ iterations of phase flip channels
$\mathrm{PF}^n$: $(a)$ $p=0.5$, $n=1$; $(b)$ $p=0.5$, $n=11$; $(c)$
$p=0.5$, $n=12$.} \label{fig:DR7}
\end{figure}

\begin{figure}[htbp]\centering
\subfigure[] {\begin{minipage}[b]{0.3\linewidth}
\includegraphics[width=1\textwidth]{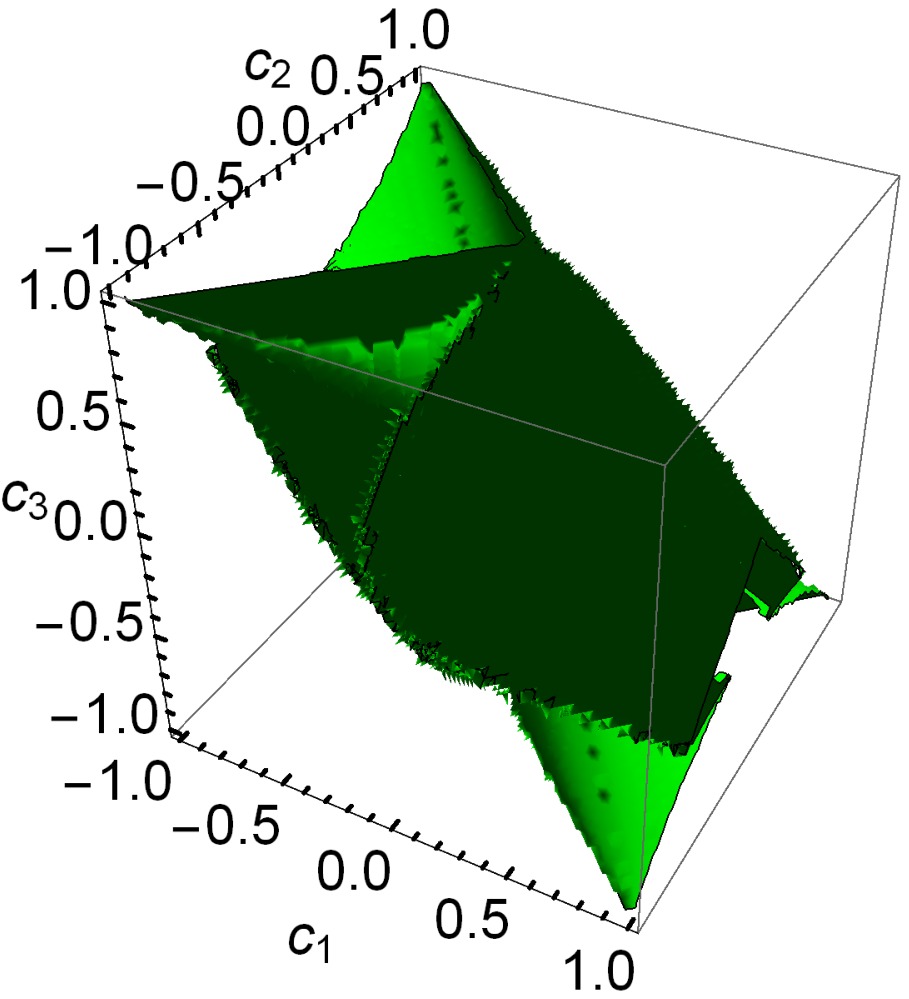}
\end{minipage}}
\subfigure[] {\begin{minipage}[b]{0.3\linewidth}
\includegraphics[width=1\textwidth]{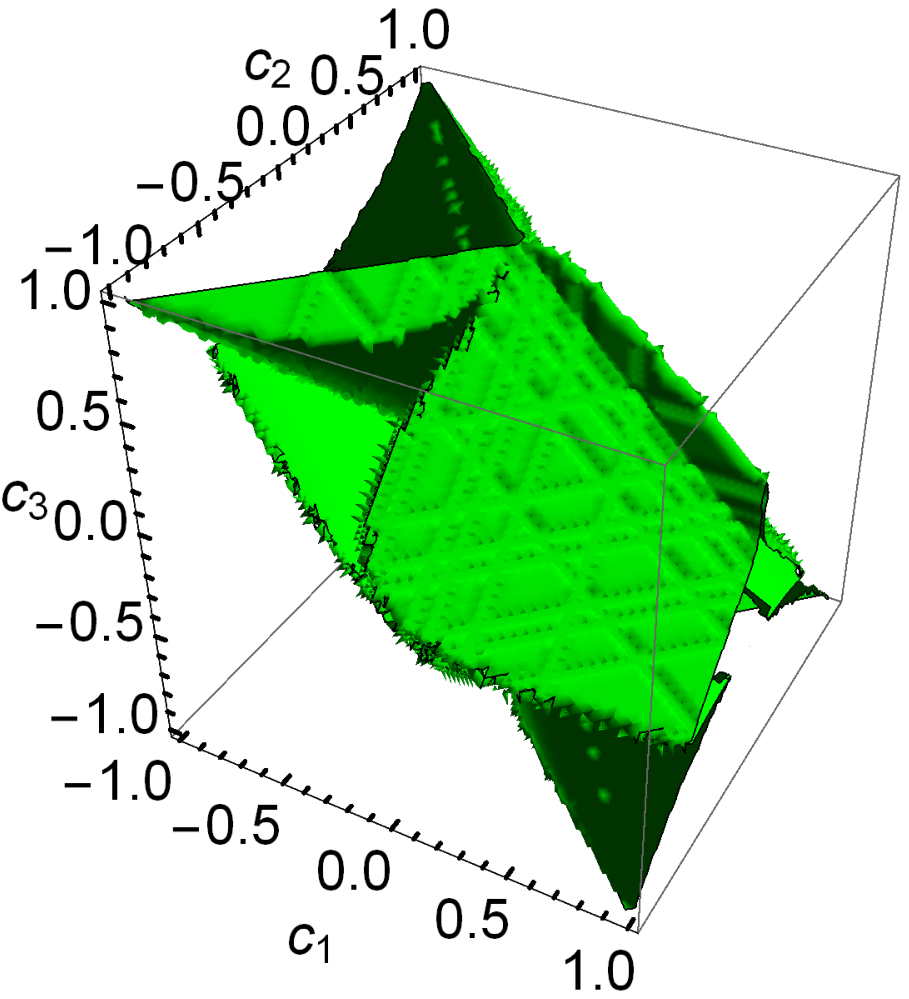}
\end{minipage}}
\subfigure[] {\begin{minipage}[b]{0.3\linewidth}
\includegraphics[width=1\textwidth]{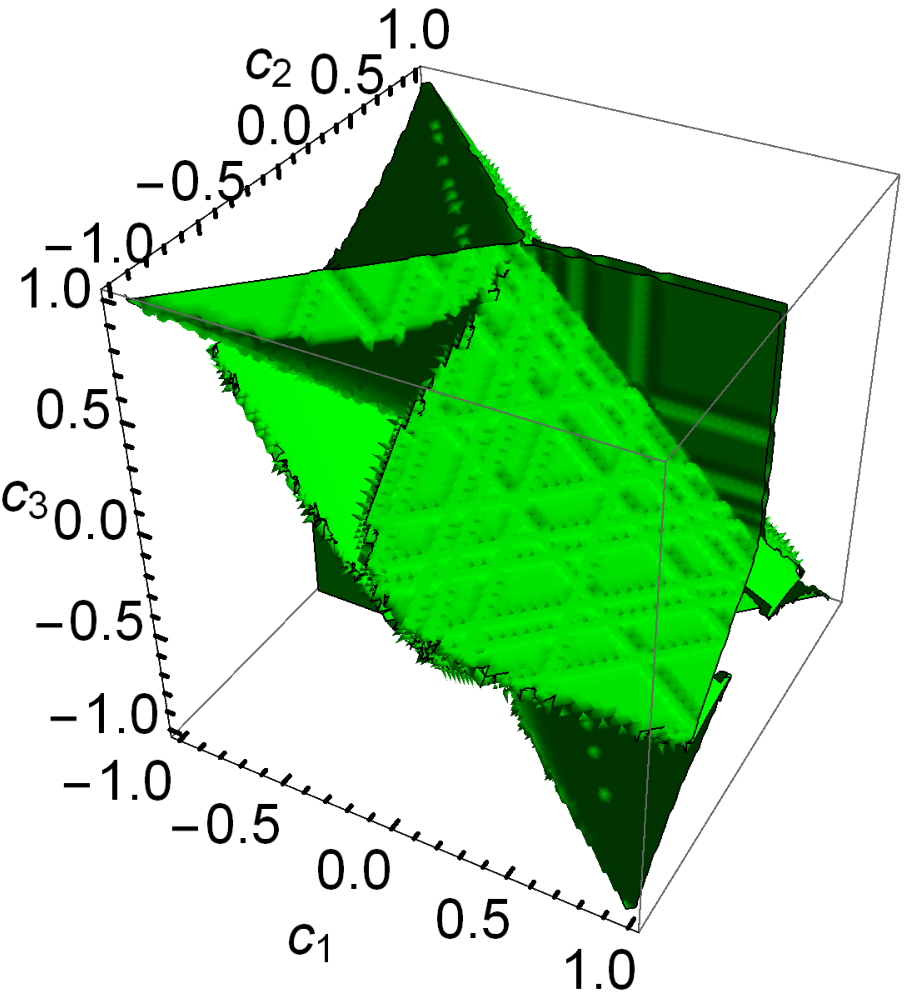}
\end{minipage}}
\caption{Surfaces of relative entropy of coherence freezing for
Bell-diagonal states under $n$ iterations of bit-phase flip channels
$\mathrm{BPF}^n$: $(a)$ $p=0.5$, $n=1$; $(b)$ $p=0.5$, $n=11$; $(c)$
$p=0.5$, $n=12$.} \label{fig:DR8}
\end{figure}

\begin{figure}[htbp]\centering
\subfigure[] {\begin{minipage}[b]{0.3\linewidth}
\includegraphics[width=1\textwidth]{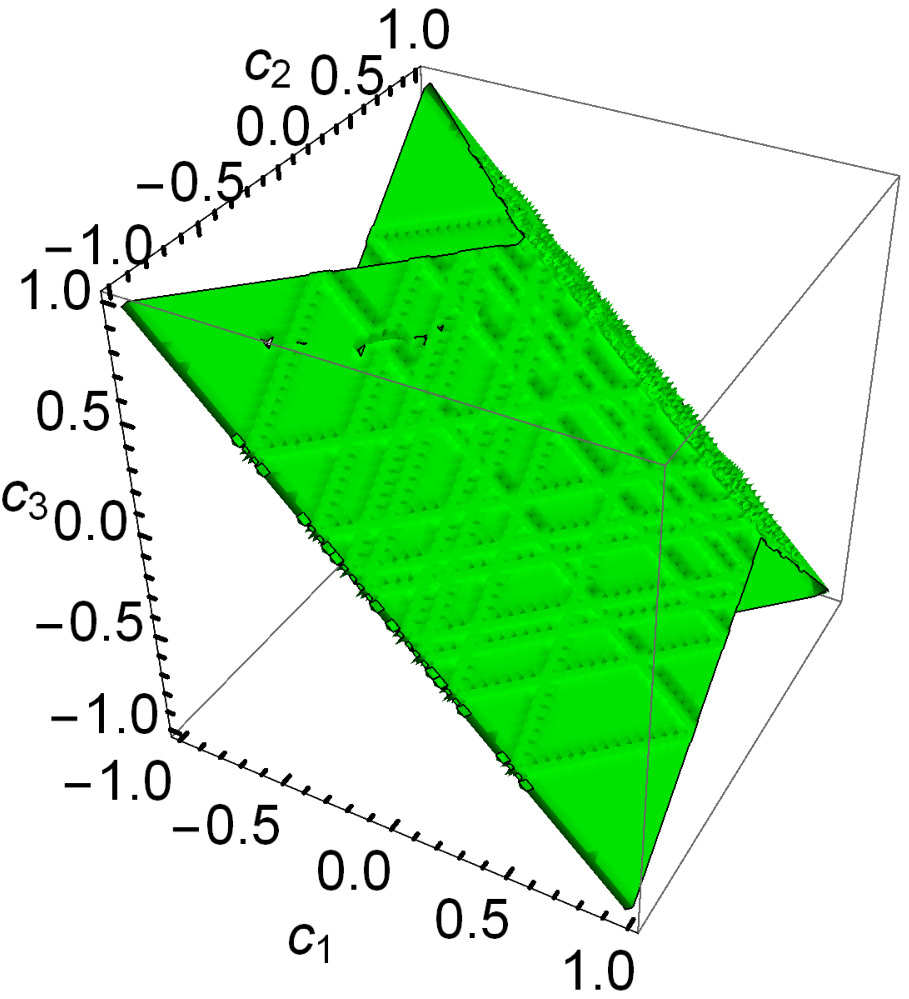}
\end{minipage}}
\subfigure[] {\begin{minipage}[b]{0.3\linewidth}
\includegraphics[width=1\textwidth]{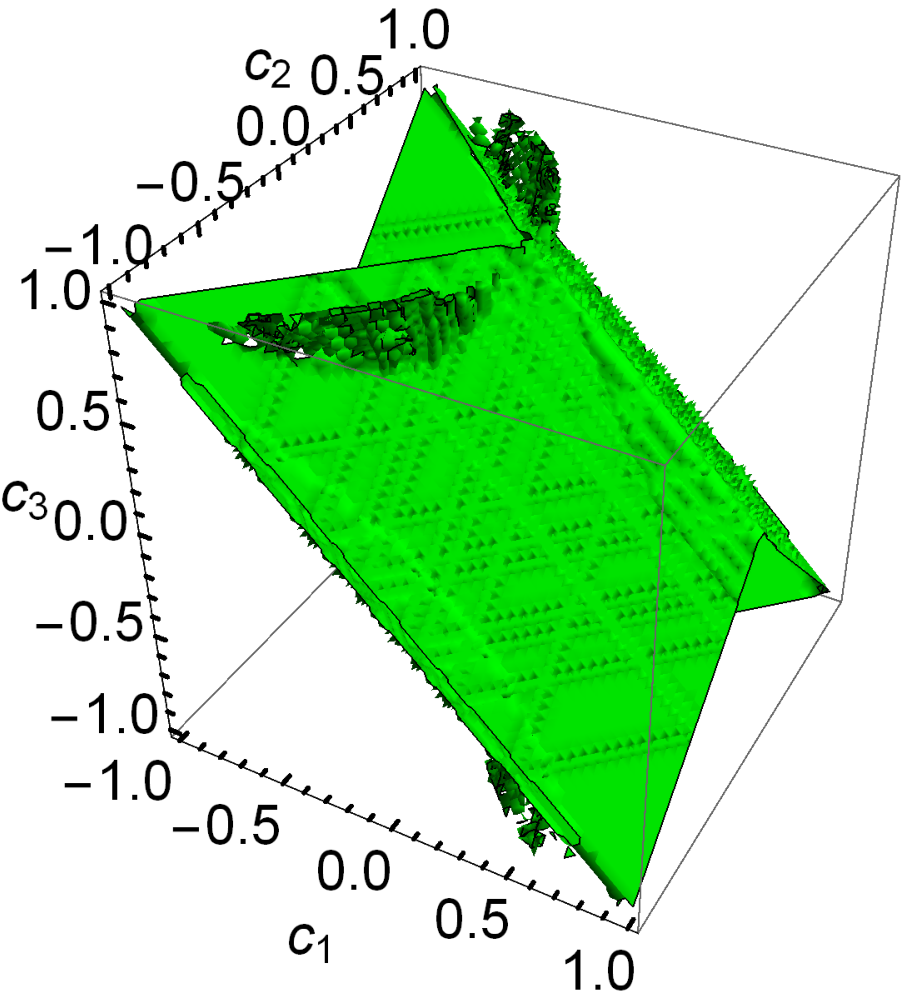}
\end{minipage}}
\subfigure[] {\begin{minipage}[b]{0.3\linewidth}
\includegraphics[width=1\textwidth]{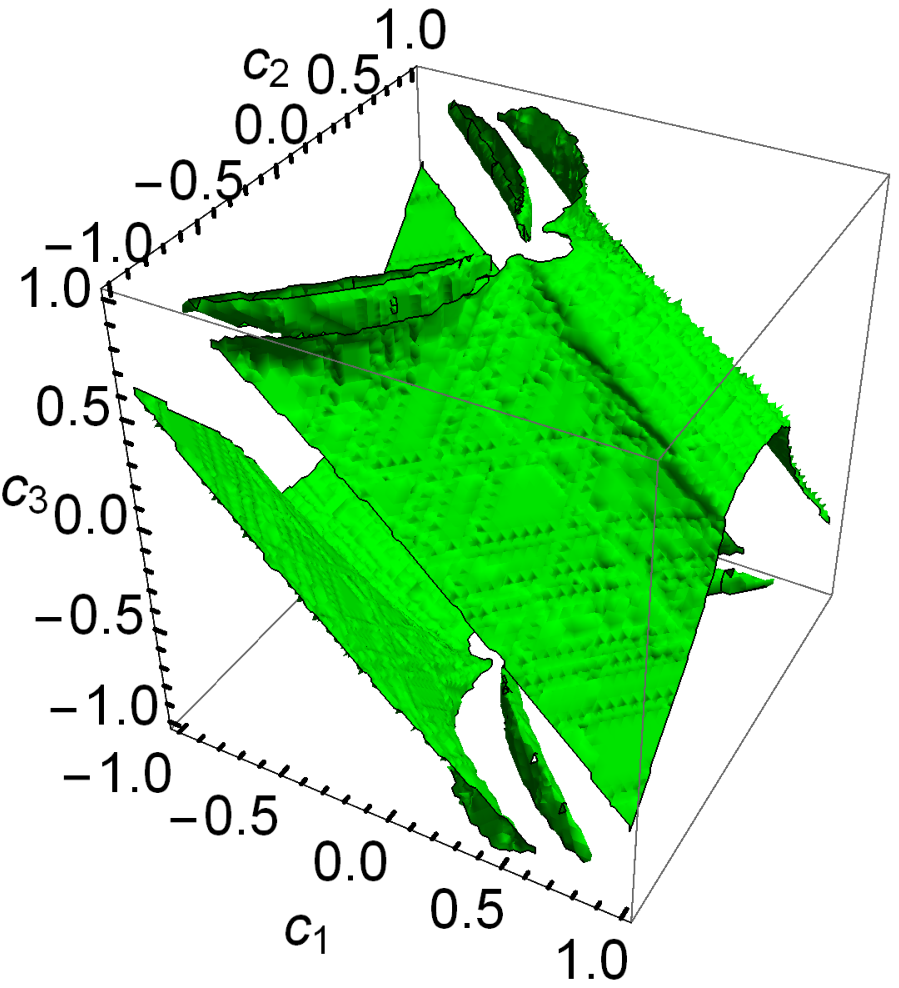}
\end{minipage}}
\caption{Surfaces of relative entropy of coherence freezing for
Bell-diagonal states under $n$ iterations of depolarizing channels
$\mathrm{DEP}^n$: $(a)$ $p=0.5$, $n=1$; $(b)$ $p=0.5$, $n=13$; $(c)$
$p=0.5$, $n=14$.} \label{fig:DR9}
\end{figure}

\begin{figure}[htbp]\centering
\subfigure[] {\begin{minipage}[b]{0.3\linewidth}
\includegraphics[width=1\textwidth]{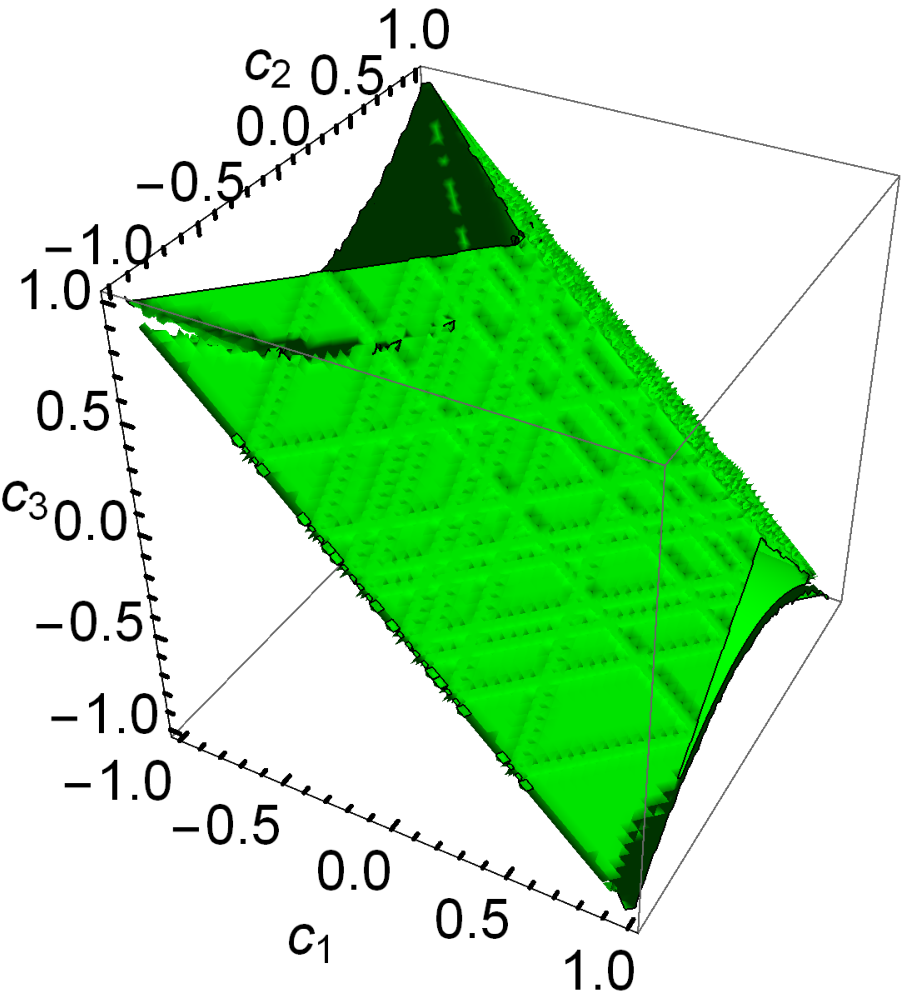}
\end{minipage}}
\subfigure[] {\begin{minipage}[b]{0.3\linewidth}
\includegraphics[width=1\textwidth]{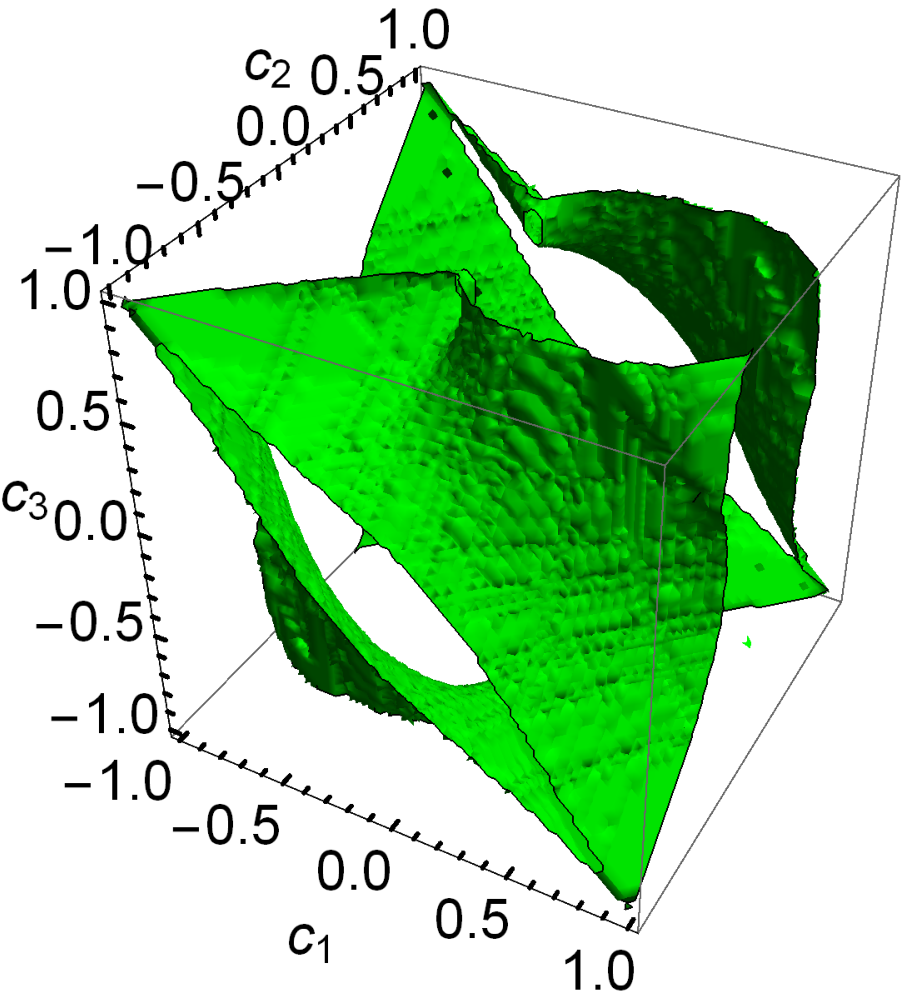}
\end{minipage}}
\subfigure[] {\begin{minipage}[b]{0.3\linewidth}
\includegraphics[width=1\textwidth]{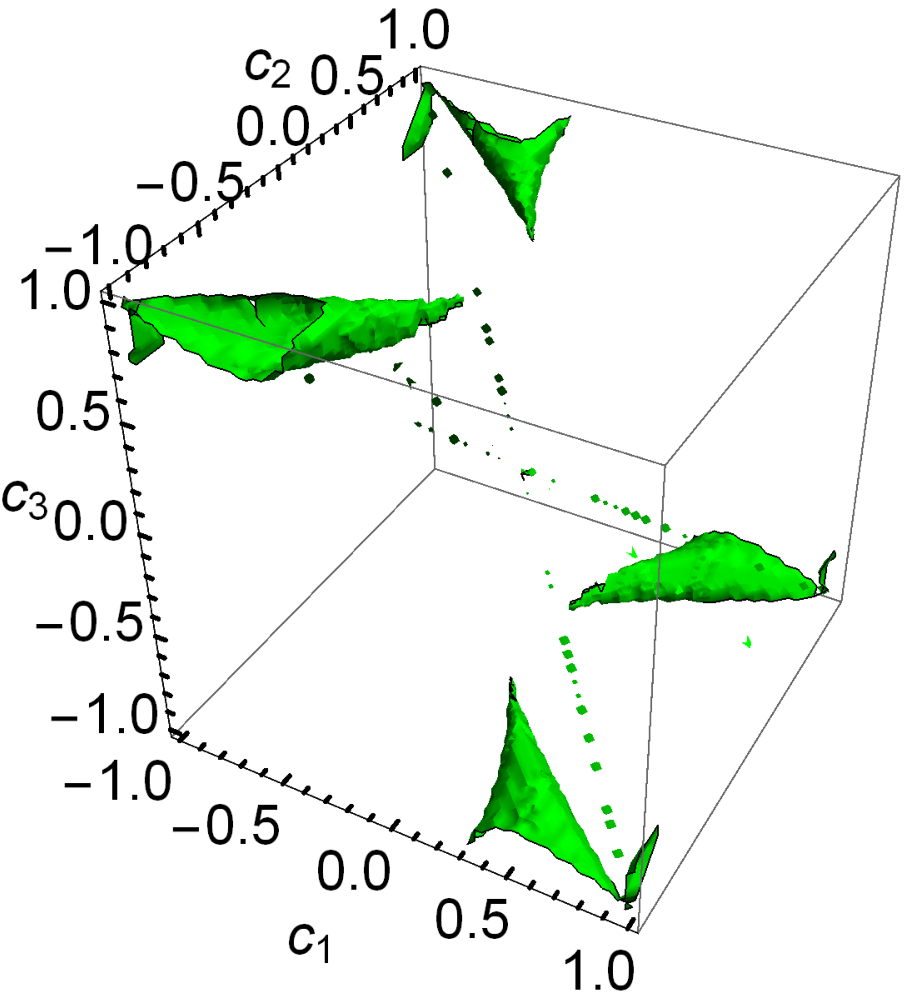}
\end{minipage}}
\caption{Surfaces of relative entropy of coherence freezing for
Bell-diagonal states under $n$ iterations of generalized amplitude
damping channels $\mathrm{GAD}^n$: $(a)$ $p=0.5$, $n=1$; $(b)$
$p=0.5$, $n=22$; $(c)$ $p=0.5$, $n=23$.} \label{fig:DR10}
\end{figure}

However, for PF, DEP and GAD channels, the situation is quite
different. For fixed $p$, when $n$ increases to a certain value, the
graph gradually disappears and becomes completely absent, i.e.,
Bell-diagonal states that satisfy such conditions do not exist. With
the increase of $p$, when $n$ is fixed the corresponding variation
in the graph will be faster under PF and GAD channels. Particularly,
for GAD channel, when $p$ is fixed, with the increase of $n$, the
pattern slowly breaks into two pieces, and further to four small
pieces distributed on the upper and lower bottom surfaces of the
cube. Then the graph becomes further smaller and disappears.

For example, for GAD channels, when $p=0.5$ and $1\leq n\leq19$, the
shape of the graphs are almost the same as Figure 10$(a)$. There is
sharp difference when $n=22$, as is shown in Figure 10$(b)$. For
$n=23$, the graph becomes four small pieces, see Figure 10$(c)$, and
disappears completely when $n=24$. For DEP channels, it seems that
with the increase of $p$, the pattern evolution accelerates. But
when $p$ is greater than 0.9, there is no such phenomena. On the
contrary, the closer $p$ is to 1, the slower the pattern evolves.

Note that when the skew information-based coherence is taken into
account, it is found that similar phenomenon occurs for coherence
freezing of Bell-diagonal states under $n$ iterations of the five
channels, while the surfaces appears to be tetrahedrons.

\vskip0.1in

\noindent {\bf 5. Conclusions}

\vskip0.1in

We have studied the $n$-th decay rates of several channels for
quantum states based on $l_{1}$ norm of coherence, relative entropy
of coherence and skew information-based coherence. The $n$-th decay
rates of non-dissipative channels for Bell-diagonal states have been
calculated and explicit formulas are derived. It has been shown that
DEP channels have the greatest impact on coherence when
$p=\frac{3}{4}$. An interesting phenomenon is that the $n$-th decay
rates of DEP channels for Bell-diagonal states decrease to zero
first and then increase. The coherence surfaces under the channels
have been plotted. We conclude that the $n$-th decay rates decrease
quickly with the increase of $n$ for fixed $p$. Moreover,
Bell-diagonal states can be completely incoherent under PF, DEP and
GAD channels, but remain coherent under BF and BPF channels.

We have also investigated the geometry of the relative entropy of
coherence and skew information-based coherence of Bell-diagonal
states under different channels when the coherence of initial state
is frozen. It has been shown that coherence freezing is quite
different for distinct channels. For fixed $n$, the evolution of the
coherence surface accelerates with the increase of $p$. For fixed
$p$, when $n$ is large enough, the freezing of initial state
coherence does not occur under PF, DEP and GAD channels, and the
states always undergo a decoherence process. For relative entropy of
coherence and skew information-based coherence, the behaviors of
coherence freezing turn out to be quite similar.

\vskip0.1in

\noindent

\subsubsection*{Acknowledgements}
This work was supported by National Natural Science Foundation of
China (Grant Nos. 11701259, 11461045, 11675113), the China
Scholarship Council (Grant No.201806825038), the Key Project of
Beijing Municipal Commission of Education (Grant No.
KZ201810028042), and Beijing Natural Science Foundation (Z190005).



\begin{thebibliography}{S2}

\bibitem{RH} Horodecki, R., Horodecki, P., Horodecki, M., Horodecki, K.: Quantum entanglement. Rev. Mod. Phys. \textbf{81}, 865 (2009)

\bibitem{HO} Ollivier, H., Zurek, W.H.: Quantum discord: a measure of the quantumness of correlations. Phys. Rev. Lett. \textbf{88}, 017901 (2001)

\bibitem{LH} Henderson, L., Vedral, V.: Classical, quantum and total correlations. J. Phys. A \textbf{34}, 6899 (2001)

\bibitem{JS} Bell, J.S.: On the problem of hidden variables in quantum mechanics. Rev. Mod. Phys. \textbf{38}, 447 (1966)

\bibitem{TB} Baumgratz, T., Cramer, M., Plenio, M.B.: Quantifying coherence. Phys. Rev. Lett. \textbf{113}, 140401 (2014)

\bibitem{SR} Rana, S., Parashar, P., Lewenstein, M.: Trace-distance measure of coherence. Phys. Rev. A \textbf{93}, 012110 (2016)

\bibitem{DG} Girolami, D.: Observable measure of quantum coherence in finite dimensional systems. Phys. Rev. Lett. \textbf{113}, 170401 (2014)

\bibitem{CN} Napoli, C., Bromley, T.R., Cianciaruso, M., Johnston, N., Adesso, G.:
Robustness of coherence: an operational and observable measure of
quantum coherence. Phys. Rev. Lett. \textbf{116}, 150502 (2016)

\bibitem{AE} Rastegin, A.E.: Quantum-coherence quantifiers based on the Tsallis relative $\alpha$ entropies. Phys. Rev. A \textbf{93}, 032136 (2016)

\bibitem{MP} Piani, M., Cianciaruso, M., Bromley, T.R.,  Napoli, C., Johnston, N., Adesso, G.:
Robustness of asymmetry and coherence of quantum states. Phys. Rev.
A \textbf{93}, 042107 (2016)

\bibitem{AW1} Winter, A., Yang, D.: Operational resource theory of coherence. Phys. Rev. Lett. \textbf{116}, 120404 (2016)

\bibitem{EC3} Chitambar, E., Streltsov, A., Rana, S.,  Bera, M.N.,  Adesso, G., Lewenstein, M.: Assisted distillation of quantum coherence. Phys. Rev. Lett. \textbf{116}, 070402 (2016)

\bibitem{EC4} Chitambar, E., Hsieh, M.H.: Relating the resource theories of entanglement and quantum coherence. Phys. Rev. Lett. \textbf{117}, 020402 (2016)

\bibitem{EC5} Chitambar, E., Gour, G.: Critical examination of incoherent operations and a physically consistent resource theory of quantum coherence. Phys. Rev. Lett. \textbf{117}, 030401 (2016)

\bibitem{CS1} Yu, C.-S., Song, H.: Bipartite concurrence and localized coherence. Phys. Rev. A \textbf{80}, 022324 (2009)

\bibitem{JM} Ma, J., Yadin, B., Girolami, D., Vedral, V., Gu, M.: Converting coherence to
quantum correlations. Phys. Rev. Lett. \textbf{116}, 160407 (2016)

\bibitem{AU} Streltsov, A., Singh, U., Dhar, H.S., Bera, M.N., Adesso, G.: Measuring quantum
coherence with entanglement. Phys. Rev. Lett. \textbf{115}, 020403
(2015)






\bibitem{LW} Wang, L., Yu, C.-S.: The roles of a quantum channel on a quantum state. Int. J. Theory. Phys. \textbf{53}, 715 (2014)

\bibitem{CS2} Yu, C.-S., Yang, S., Guo, B.: Total quantum coherence and its applications. Quantum Inf. Process. \textbf{15}, 3773 (2016)

\bibitem{VG1} Giovannetti, V., Lloyd, S., Maccone, L.: Quantum-enhanced measurements: beating the standard quantum limit. Science \textbf{306}, 1330 (2004)

\bibitem{RDD} Demkowicz-Dobrza\'nski, R.,  Maccone, L.: Using entanglement against noise in quantum metrology. Phys. Rev. Lett.  \textbf{113}, 250801 (2014)

\bibitem{VG2} Giovannetti, V.,  Lloyd, S.,  Maccone, L.: Quantum metrology. Phys. Rev. Lett. \textbf{96},  010401 (2006)

\bibitem{RJ} Glauber, R.J.: Coherent and incoherent states of the radiation field. Phys. Rev. \textbf{131}, 2766 (1963)

\bibitem{EC1} Sudarshan, E.C.G.: Equivalence of semiclassical and quantum mechanical descriptions of statistical light beams. Phys. Rev. Lett. \textbf{10}, 277 (1963)

\bibitem{LM} Mandel, L., Wolf, E.: Optical Coherence and Quantum Optics, Cambridge University Press, Cambridge (1995)

\bibitem{GS} Engel, G.S., Calhoun, T.R., Read, E.L., Ahn, T.-K., Mancal, T., Cheng, Y.-C., Blankenship, R.E., Fleming, G.R.: Evidence for wavelike energy transfer through quantum coherence in photosynthetic
systems. Nature (London) \textbf{446}, 782 (2007)

\bibitem{MB} Plenio, M.B., Huelga, S.F.: Dephasing-assisted transport: quantum networks and biomolecules. New J. Phys. \textbf{10}, 113019 (2008)

\bibitem{EC2} Collini, E., Wong, C.Y., Wilk, K.E., Curmi, P.M.G., Brumer, P., Scholes, G.D.: Coherently wired light-harvesting in photosynthetic marine algae at ambient temperature. Nature (London) \textbf{463}, 644 (2010)

\bibitem{SL} Lloyd, S.: Quantum coherence in biological systems. J. Phys. Conf. Ser. \textbf{302}, 012037 (2011)

\bibitem{CML} Li, C., Lambert, N., Chen, Y., Chen, G., Nori, F.: Witnessing quantum coherence: from solid-state to biological systems. Sci. Rep. \textbf{2}, 885 (2012)







\bibitem{SH} Huelga, S., Plenio, M.: Vibrations, quanta and biology. Contemp. Phys. \textbf{54}, 181 (2013)




\bibitem{MDL} Lang, M.D., Caves, C.M.: Quantum discord and the geometry of Bell-diagonal states. Phys. Rev. Lett. \textbf{105}, 150501 (2010)

\bibitem{BL} Li, B., Wang, Z.-X., Fei, S.-M.: Quantum discord and geometry for a class of two-qubit states. Phys. Rev. A \textbf{83}, 022321 (2011)

\bibitem{YKW1} Wang, Y.-K., Ma, T., Fan, H., Fei, S.-M.,
Wang,Z.-X.: Super-quantum correlation and geometry for Bell-diagonal
states with weak measurements. Quantum Inf. Process. \textbf{13},
283 (2014)

\bibitem{YKW2} Wang, Y.-K., Shao, L.-H.,  Ge, L.-Z., Fei, S.-M.  Wang, Z.-X.: Geometry of Quantum Coherence for Two Qubit X States. Int. J. Theor. Phys.
\textbf{58}, 23 (2019)

\bibitem{YKW3} Wang, Y.-K.,  Fei, S.-M., Wang, Z.-X.: Dynamics of quantum coherence in Bell-diagonal states under Markovian channels. Commun. Theor. Phys. \textbf{71}, 555 (2019)






\bibitem{TRB1} Bromley, T.R., Cianciaruso, M., Adesso,G.: Frozen quantum coherence. Phys. Rev. Lett. \textbf{114}, 210401 (2015)

\bibitem{XDY} Yu, X., Zhang, D.,  Liu, C.,  Tong, D.: Measure-independent freezing of quantum coherence. Phys. Rev. A \textbf{93}, 060303 (2016)



\bibitem{CLL} Liu, C.,  Yu, X., Xu, G., Tong, D.: Ordering states with coherence measures. Quantum. Inf. Process \textbf{15}, 4189-4201 (2016)

\bibitem{AW2} Winter, A., Yang, D.: Operational resource theory of coherence. Phys. Rev. Lett. \textbf{116}, 120404 (2016)

\bibitem{DM}  Gao, D., Xin, Q., Ye, Z., He, C., Qiu, J.: Decay rate of the $l_1$ norm coherence in single-qubit
systems. Int. J. Theor. Phys. \textbf{58}, 1568-1575 (2019)

\bibitem{MAN} Nielsen, M.A., Chuang, I.L.: Quantum Computation and Quantum Information, Cambridge University Press,
Cambridge (2000)

\bibitem{CS3} Yu, C.-S.: Quantum coherence via skew information and its polygamy. Phys. Rev. A \textbf{95}, 042337 (2017)





\bibitem{MPSS} Montealegre, J.D., Paula, F.M., Saguia, A., Sarandy, M.S.: One-norm geometric quantum discord under decoherence. Phys. Rev. A \textbf{87}, 042115 (2013)



\vskip0.2in


\end{thebibliography}
\end{document}